\documentclass[12pt,onecolumn]{aastex63}
\usepackage{bm}
\usepackage{ulem}

\shortauthors{Kino et al.}
\shorttitle{}

\begin{document}

\title{
Morphological transition of the
compact radio lobe in 3C84
via the strong jet-cloud collision}

\correspondingauthor{Motoki Kino}
\email{motoki.kino@nao.ac.jp}

\author[0000-0002-2709-7338]{Motoki Kino}
\affil{Kogakuin University of Technology \& Engineering, Academic Support Center, 
 2665-1 Nakano, Hachioji, Tokyo 192-0015, Japan}
\affil{National Astronomical Observatory of Japan, 2-21-1 Osawa, Mitaka, Tokyo, 181-8588, Japan}

\author[0000-0002-8169-3579]{Kotaro Niinuma}
\affil{Graduate School of Sciences and Technology for Innovation, Yamaguchi University, 1677-1 Yoshida, Yamaguchi, Yamaguchi 753-8512, Japan}

\author[0000-0003-2535-5513]{Nozomu Kawakatu}
\affil{National Institute of Technology, Kure College, 2-2-11, Agaminami, Kure, Hiroshima, 737-8506, Japan}

\author[0000-0003-0292-3645]{Hiroshi Nagai}
\affil{National Astronomical Observatory of Japan, 2-21-1 Osawa, Mitaka, Tokyo, 181-8588, Japan}
\affil{Department of Astronomical Science, The Graduate University for Advanced Studies, SOKENDAI, 2-21-1 Osawa, Mitaka, Tokyo
181-8588, Japan}

\author[0000-0003-4916-6362]{Gabriele Giovannini}
\affil{INAF - Istituto di Radioastronomia, via Gobetti 101, 40129 Bologna, Italy}
\affil{Dipartimento di Fisica e Astronomia, Universita' di Bologna, via Gobetti 103/2, I-40129, Bologna, Italy}

\author[0000-0003-4470-7094]{Monica Orienti}
\affil{INAF - Istituto di Radioastronomia, via Gobetti 101, 40129 Bologna, Italy}

\author[0000-0003-3823-7954]{Kiyoaki Wajima}
\affil{Korea Astronomy and Space Science Institute, 776 Daedeokdae-ro, Yuseong-gu, Daejeon 34055, Republic of Korea}
\affil{Department of Astronomy and Space Science, University of Science and Technology, 217 Gajeong-ro, Yuseong-gu, Daejeon 34113, Republic of Korea}

\author[0000-0001-7618-7527]{Filippo D'Ammando}
\affil{INAF - Istituto di Radioastronomia, via Gobetti 101, 40129 Bologna, Italy}

\author[0000-0001-6906-772X]{Kazuhiro Hada}
\affil{Mizusawa VLBI Observatory, National Astronomical Observatory of Japan, 2-21-1 Osawa, Mitaka, Tokyo 181-8588, Japan}
\affil{Department of Astronomical Science, The Graduate University for Advanced Studies, SOKENDAI, 2-21-1 Osawa, Mitaka, Tokyo
181-8588, Japan}

\author[0000-0002-8657-8852]{Marcello Giroletti}
\affil{INAF - Istituto di Radioastronomia, via Gobetti 101, 40129 Bologna, Italy}

\author[0000-0003-0685-3621]{Mark Gurwell}
\affil{Center for Astrophysics | Harvard \& Smithsonian, 
60 Garden Street, Cambridge, MA 02138, USA}

\begin{abstract}

We report multi-epoch Very Long Baseline Interferometric (VLBI) observations of the compact radio lobe 
in the radio galaxy 3C84 (NGC1275) during 2016 - 2020. 
%%%%%%
The image sequence of 3C84 reveals that the hotspot 
in the radio lobe showed the one-year long frustration 
in 2017 within a compact region of about 0.07 parsec, 
suggesting a strong collision between the jet and  a compact dense cloud with the estimated average density 
about $(4\-- 6)\times 10^{5}~{\rm cm^{-3}}$.
Although the hotspot and the radio lobe began to move south again after its breakout, 
the radio lobe showed a morphological transition
from FR II- to FR I-class radio lobe 
and its radio flux became fainter. 
This is the first detection of 
the dynamical feedback from the cloud to the jet 
where the cloud located on the jet axis significantly
interferes with the jet propagation and evolution 
at the central one-parsec region in
active galactic nucleus.

\end{abstract}

\section{Introduction}
\label{sec:intro}

Radio jets in active galactic nuclei (AGNs) are considered to be one of the major drivers of galaxy evolution
and understanding how jets carry energy from the central nucleus to large scales is one of the most fundamental questions
\citep[e.g.,][for review]{Fabian12, kormendy13}. 
Radio lobes associated with the jets are 
key structures reflecting the history of the jet propagation 
and its radio-mode feedback to the surrounding matter
\citep[e.g.,][]{Croton06,Wagner12}.
The radio lobes are classified into two classes.
While Fanaroff-Riley class I (FR I) are core-brightened with fainter outer edge of the radio lobes, Fanaroff-Riley class II (FR II) are edge-brightened with bright hotspots at the ends of their lobes \citep[][]{fanaroff74}. 
%%%%%
%FRII jets are likely relativistic still at large scales, 
%as indicated by the brightness asymmetry between jet and %counter-jet (e.g., Cyg A, Carilli \& Barthel 1996), 
%while FRI jets (e.g., 3C 31, Laing \& Bridle 2002) 
%seem to be efficiently decelerated on kpc scale
%from the central engine (see, e.g. Bicknell 1984; 
%Laing 1993, 1996; Laing \& Bridle 2014).
%%%%%%%%%
There are two scenarios for the origin of the FR~I/FR~II divide.
One is due to a different nature of the central engine that 
drives the jet
\citep[e.g.,][]{rawlings91,baum95,Meier97,ghisellini01},
while the other is due to a different degree of deceleration caused by intergalactic matter
\citep[e.g.,][]{Kawakatu08,kharb12,laing14}. 
%%%%%%%%%%%%%%
% motivation
%%%%%%%%%%%%%%
The purpose of this study is to 
explore energy transport by radio lobes on parsec scale 
that has yet to be studied sufficiently.

The nearby radio galaxy 3C84 ($z=0.0176$) at the center of the Perseus cluster, harboring a supermassive black hole with its mass of 
$M_{\rm BH}=(0.8-2)\times 10^{9}M_{\odot}$
\citep[][]{scharwarchter13, Giovannini18}
is an excellent laboratory for exploring the physics of energy transport by radio lobes at  parsec scale. 
%%%
The C3 component in 3C84 is known as a newborn radio lobe component associated with the radio outburst that started in 2005
and it propagates southward 
and becomes brighter \citep{Nagai10, Hiura18}.
The brightness peak in the C3 component is identified 
as the hotspot, 
which is a termination shock at the tip of the jet.
%%%
The peculiar structure of C3 suggests interactions between the jet and the surrounding medium \citep{Giovannini18} 
and a recent VLBI observation 
found a positional flip of the hotspot in 2015 September that indicated 
the jet and cloud interaction
\citep{Kino18}. 
VLBA data also showed an enhancement of linearly polarized flux 
which also supports the jet-cloud interaction \citep{Nagai17}.
In this Letter, we report continuous VLBI monitoring
observations of 3C84 that reveals further strong jet-cloud collision
that accompanying 
morphological transition of the radio lobe.
%%%%%
With the cosmology parameters of 
$\Omega_{m}=0.27$, $\Omega_{\Lambda}=0.73$, and 
$H_{0}= 71~\rm{km~s^{-1}~Mpc^{-1}}$ 
\citep[][]{komatsu09}
the angular scale of 1 mas corresponds to 
a linear distance of 0.35 parsec for 3C84.

%at 22~GHz and  43~GHz, with the combined VLBI array of Korean VLBI Network (KVN) and VLBI 
%Exploration of Radio Astrometry (VERA), named KaVA.
%In addition to our KaVA data, we also used archived
%Very Long Baseline Array (VLBA) data for further data enhancement.
%%%%%%%
%%%%%%%
%In this Letter, we will report our witnessing of 
%ongoing radio-mode feedback 

\section{Observations and data reductions}
\label{sec:obs-data}
%%%%%%%%%%%%%%%%%%%
% KaVA 43GHz Data.
%%%%%%%%%%%%%%%%%%%
We have conducted the high-cadence monitoring 
observations of 3C84 with the KaVA array at 43~GHz during 2016 - 2018. 
%%%
%The KaVA Array is operated by Korea Astronomy and Space Science Institute (KASI) and the National Astronomical Observatory of Japan (NAOJ). 
%All the data shown here were recorded at 1 Gbps (256 MHz bandwidth, %16MHz $\times$ 16 subbands) 
Imaging capabilities of KaVA are summarized 
in \citep{Niinuma14}. 
%%%
%The on-source time for 3C84 was about 6-7 hr, which provided
%good uv coverages. 
%%%
The typical size of the original beam of KaVA at 43 GHz is 0.6-0.7~mas
\citep[see also][]{Kino18, Wajima20}.
%%%
%and we restored all of the images with a beam size of 0.75 mas}% 
%%
%%%%%%%%%%%%%%%%%%%%%%%%%
% VERA Genji 22GHz Data.
%%%%%%%%%%%%%%%%%%%%%%%%%
At 22~GHz, we have conducted another high-cadence monitoring observation with the VERA array in the framework of the GENJI programme (Gamma-ray Emitting Notable AGN Monitoring with Japanese VLBI).
%%%%
%aims for dense sampling of gamma-ray loud AGNs using the available calibrator time 
%in the Galactic maser astrometry project of VERA. 
%%%
%Maser sessions need to monitor a bright calibrator once in every 80 minutes, for which we use GENJI sources including 3C84. 
%%%
%During each observation, on-source time for 3C 84 was 
%typically 30 minutes, consisting of 4-6 scans at different 
%hour angles. 
%%%
Data reduction was performed using the National Radio Astronomy Observatory (NRAO) Astronomical Imaging
Processing System (AIPS) in the same way described in GENJI programme \citep{Nagai13,Hiura18}.
%%%%%%
%The final images were obtained after a number of iterations with modelfit and self-calibration implemented in the Difmap.
%%%%%%%
A priori amplitude calibration was applied using the measured system noise
temperature and the elevation-gain curve of each antenna. 
We calibrated the bandpass characteristics of phase and amplitude at each station using the auto-correlation data. 
Following the amplitude calibration, fringe-fitting was performed to calibrate the visibility phases. 
Finally, the data were averaged over each
intermediate frequency band. The imaging and self-calibration were performed in the Difmap software 
%\citep{Shepherd97} 
in the usual manner.
%%%%%%%%%%%%%%
% BU 43GHz Data.
%%%%%%%%%%%%%%
We also included VLBA 43GHz archived data obtained in the  
blazar monitoring program led by Boston University group
(https://www.bu.edu/blazars/).
%%%%
%which totally support and guarantee our observational 
%finding obtained by KaVA and VERA. 
%%%%
To determine the brightness peak position of the C3 component, 
we performed two-dimensional Gaussian fit to the visibility data by using Difmap task modelfit. 
In the process, we trimmed off VLBA visibility data on the baselines longer than $350~\rm{M\lambda}$ to align with KaVA's $uv$-coverage.
%%%
%to discuss the overall structures and fluxes with a 
%consistent spatial scale. 
%%%%%%%%%%%%%%%%%%%
% SMA observations
%%%%%%%%%%%%%%%%%%%
For better understanding of the light curve behavior,
we included the Submillimeter Array (SMA)  monitoring data
at 1.3~mm.
Detailed explanations for the monitoring program 
are described in the literature \citep{Gurwell07}.

\section{Results}
\label{sec:results}

%Here, we report our high cadence VLBI observations of 3C84 that indicates a strong jet-cloud collision in 2017 and a subsequent drastic change of the radio lobe morphology.

\subsection{Discovery of the frustrated hotspot}

\begin{figure}[t]
\begin{center}
\includegraphics[trim=0mm 0mm 0mm 0mm, clip, width = 180mm]
{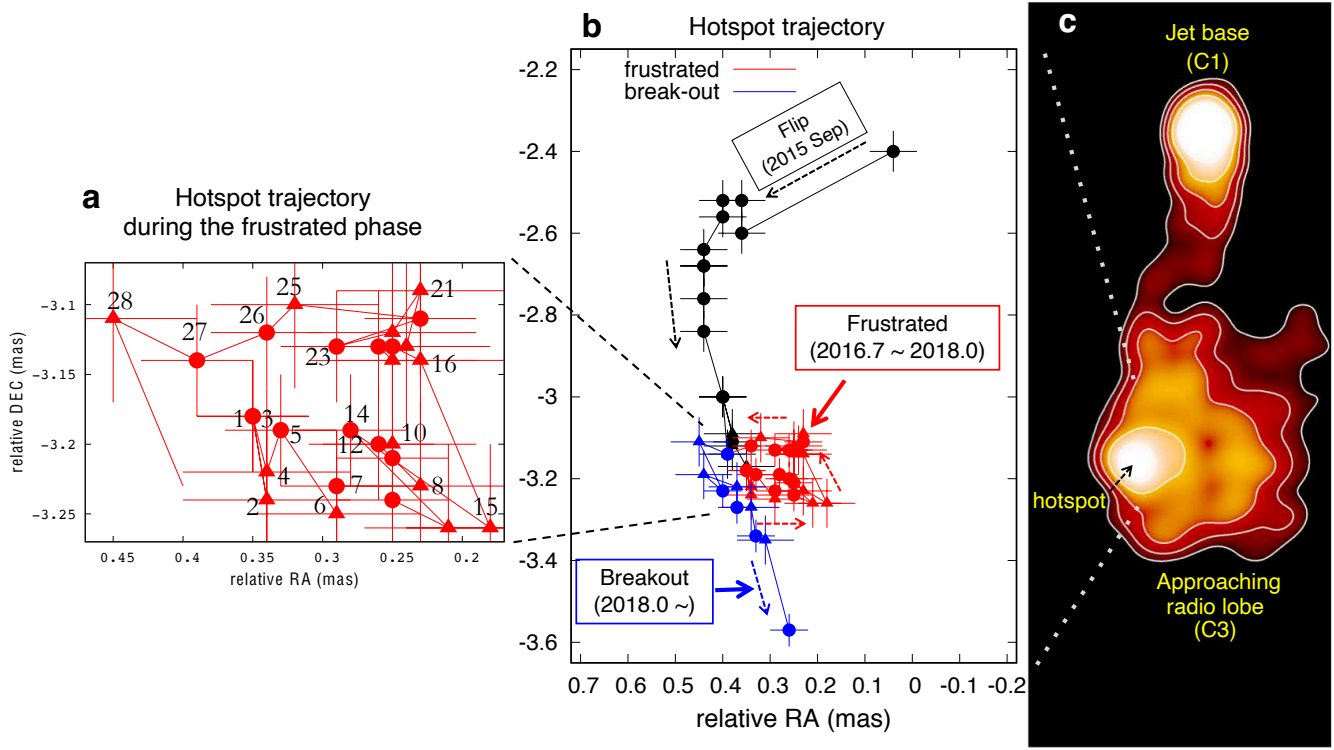}
%%%%%
%\includegraphics[width = 120mm]{vortex.eps}
%%%%%
\caption{\small\textbf{Trajectory of the hotspot in C3.} 
%%%%%%%%%%%
\textbf{a,}
%%%%%%%%%%%
Zoomed-in view of the hotspot (the brightness peak position in C3) 
trajectory during the frustrated phase.
The number next to each position data indicates the epoch number (28 epochs in total).
During the frustrated phase, the positions of the 
hotspot were not random, 
but they showed a counterclockwise trajectory.
The circle dots correspond the data points measured by KaVA,
while the triangle dots are those measured by VLBA.
%%%%%%%%%%%
\textbf{b,}
%%%%%%%%%%%
Overall sequence of the brightness peak position change the hotspot component. 
The arrows indicate the time sequence of the position.
The flip of the hotspot was found in 2015 September
\citep{Kino18}.
%%%%
As shown in the panel a, the position of the hotspot 
showed the counter-rotating trajectory 
in the sky plane in the compact region with its projected angular size 
about 0.2 mas (red dots).
After the frustrated phase, the C3 resumed to 
propagate moving south (blue points).
%%%%%%%%%%%
\textbf{c,}
%%%%%%%%%%%
The entire image of the 3C84 jet with VLBA at 43~GHz.
The C1 component corresponds to the jet base 
containing the central engine that drives the jet.
%%%%
%Jets that interact with surrounding matter and are 
%slowed down have an extended structure and are 
%called radio lobes. In this figure, we call it as the jet head
%and the C3 component corresponds to the jet head.
%%%%%%%%%
%Bottom panel:
%%%%%%%%%
%A sketch of the jet head propagation in the dense cloud in a 3D view. 
%A 3D helical motion can naturally reproduce a
%2D rotating motion when the jet viewing angle 
%is simply deviated from 90 degree.
%%%
%When it breakouts the dense cloud, the C3' component emerges
%and actual images of the emergence of C3' will be
%shown in Figure~\ref{fig:breakout}.
\label{fig:C3trajectory}}
\end{center}
\end{figure}

%%%%%%%%%%%%%%%%%%%%%
%  C3 trajectory
%%%%%%%%%%%%%%%%%%%%%%
In Figure~\ref{fig:C3trajectory}, 
we present the time sequence of the positional change of the hotspot
obtained from  more than 50 epochs of images observed 
with KaVA and VLBA at 43~GHz. 
One can readily see new important findings as follows:
(1) 
As shown in the panel b of Figure~\ref{fig:C3trajectory},
the position of the hotspot was frustrated during the period 
from 2016.7 to 2018.0. The location of the hotspot during 
the frustrated phase was confined within a compact region 
with its angular size about $0.2$~mas, which
is equivalent to about 0.07~pc in the projected physical scale.
%%%%%%%%%%%
(2) 
The trajectory of the hotspot during the frustrated phase 
was not random but rather smoothly connected between neighboring epochs and 
it showed an apparent counter-clockwise rotating motion on the sky plane 
(shown in the panel a of Figure~\ref{fig:C3trajectory}).
The trajectory of the hotspot may 
remind us of a similarity with 
the dentist-drill model proposed by \citet[][]{scheuer82} 
where hotspots show vivid wobbling.
(3)
After 2018.0, it broke out and began to move south again. 
%%%%
These frustration and breakout of the hotspot 
cannot be explained by a global change of the entire jet 
direction due to the precession of 
the jet base \citep[][]{Hiura18},
while a strong collision between the jet and
a compact dense cloud at this local site 
can naturally account for them.
%%%%%
%In the lower panel of Figure~\ref{fig:C3trajectory}, we add a sketch explaining the  expected phenomenon.
%%%%%%%%%%%

%In addition, 
%the trajectory of the C3 should not be confined 
%in two dimensional (2D) rotating motion 
%but a three dimensional (3D) one.
%It is natural to suppose that 
%a (partial) 3D helical motion reproduces 
%the apparent 2D rotating motion in the sky plane.
%%%%%%%%%%%%%
%In the lower panel of Figure~\ref{fig:C3trajectory},
%We show a supplemental sketch explaining this geometrical effect.
%%
%Since the viewing angle of the 3C84 jet is estimated as 
%$25^{\circ} \lesssim \theta_{\rm view} \lesssim 49^{\circ}$,
%the projection effect just shorten the observed length by the factor %of 1.3-2.4 from the real length.

\begin{figure}[t]
\includegraphics
[width=18cm]
{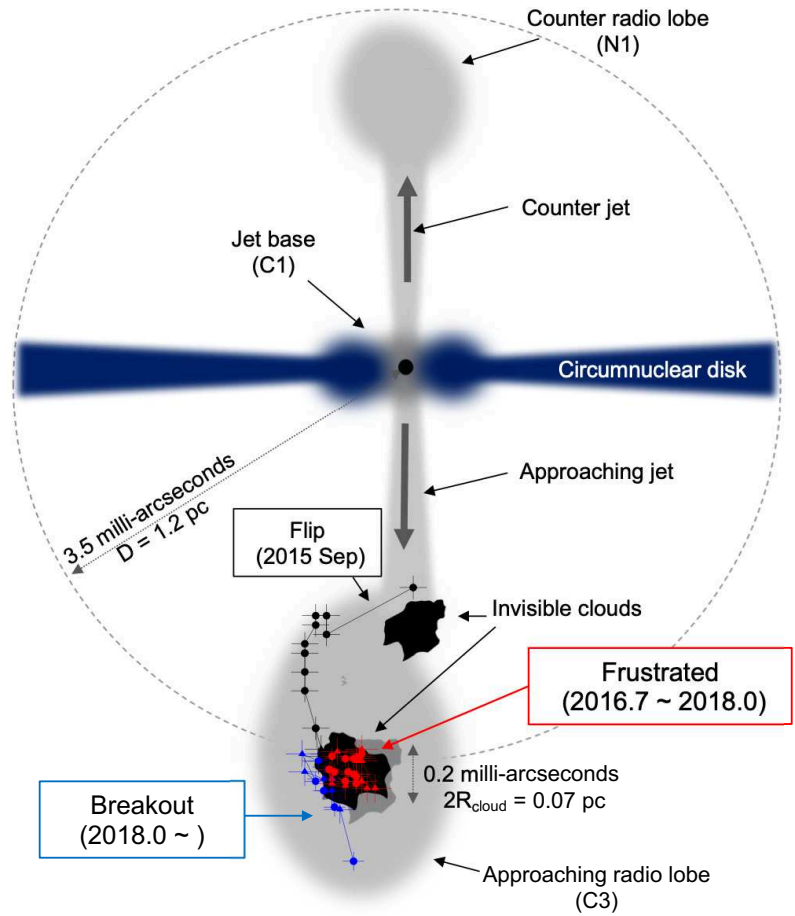}
\caption
{
%%%%%%%%%%%
\textbf{a,}
%%%%%%%%%%%
The observed trajectory of the hotspot overlaid onto a schematic picture of the central parsec scale region of 3C84.
Before the frustrated phase (the black dots),
the C3 component continued to propagate towards south. 
%%%
%and this part has been already shown in 
%the previous report \citep{Kino18}.
%%%
In the period from late 2016 to the end of 2017, 
we discovered  1.3 years of
frustration of the hotspot in C3 (red dots). 
%%%
%In the frustrated phase, 
%it is found that the hotspot in
%the C3 component suddenly stopped its propagation.
%%%
%At closer look into this frustrated phase,
%we identify a small bouncing motion along the north-south direction.
%%%
For completeness, we add an illustration of 
the counter lobe component (N1), whose radio emission 
is largely interrupted by free-free absorption caused 
by the foreground circumnuclear disk \citep{Walker00, Wajima20}.
%%%%%%%%%%%%
%\textbf{b,}
%%%%%%%%%%%
%The corresponding time-sequence
%of the positional change of the C3 component in Declination is shown.
}
\label{fig:overall}
\end{figure}

%%%%%%%%%%%%%%%%%%%
% overall picture
%%%%%%%%%%%%%%%%%%%%
In Figure~\ref{fig:overall}, 
we summarize the overall picture of the
hotspot trajectory relative to the peak position 
of the jet base (hereafter C1). 
The projected distance from the jet base (C1) to 
C3 was about one-parsec scale during our observational period.
One can see the overall trajectory of the hotspot toward the south.
In the beginning, the abrupt large flip of the hotspot position 
was found in 2015 August-September and it was explained 
an off-axis collision between the jet and a dense cloud \citep{Kino18} .
After the flip in 2015, the hotspot propagated towards south until the middle of 2016. 
Then, the hotspot suddenly stopped its propagation
over the period from 2016.7 (4th epoch in the red dot data)
to 2018.0 (i.e., $t_{\rm frust} \approx 1.3$~years).
%%%
Such a phenomenon has never been found in AGN jets.
%%%%%%%%%%%%%%%%%%%%
% basic energetics
%%%%%%%%%%%%%%%%%%%%
The strong head-on collision releases 
the bulk kinetic energy of the 3C84 jet into other forms of energy.
 The amount of the released energy by the collision is 
 one of the most interesting quantities,
 and it can be estimated as 
 $E_{\rm feedback}\approx L_{\rm jet}t_{\rm frust}$.
%%
%The amount of the liberated energy by the collision is one of the most interesting quantities
%and it can be straightforwardly estimated.
%%
Since the duration of the frustration and the size of 
the cloud are directly determined from the kinematics of the hotspot, 
we obtain $E_{\rm feedback}\approx 4 \times 10^{52-53}~{\rm erg}$
where the value of the total jet kinetic power of 3C84 is
$L_{\rm jet}\approx 1\times 10^{45-46}~{\rm erg~s^{-1}}$ 
suggested by the dynamics of X-ray cavity in the Perseus cluster
inflated by the jet kinetic power \citep{Heinz98, churazov00}.
%
%Although it is not trivial how $E_{\rm feedback}$ is consumed 
%from the observation data during the frustrated phase alone, 
%the subsequent observational data after 2018 to be shown 
%%%
In \S~\ref{sec:discussion},
we will discuss possible usages of $E_{\rm feedback}$.

\subsection{Light curves}

\begin{figure}[t]
\begin{center}
\includegraphics[trim=0mm 0mm 0mm 0mm, clip, width = 180mm]
{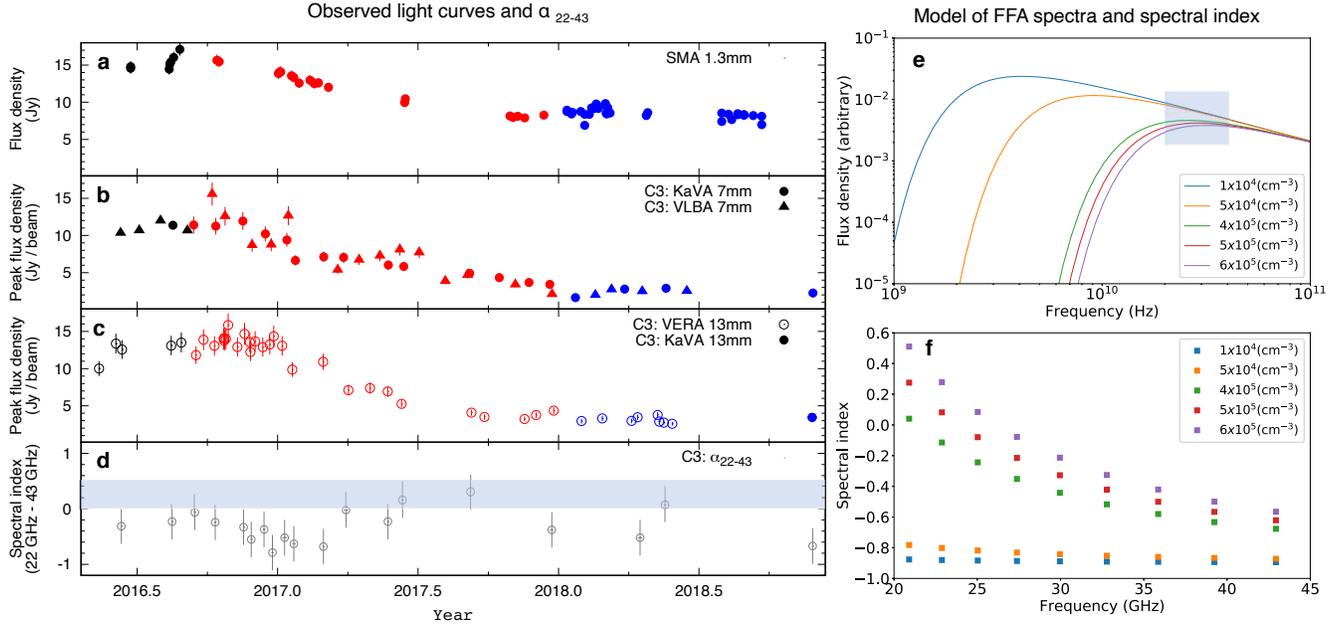}
\caption{Left: Observed light curves of the C3 component
from 22GHz (13~mm) to 230GHz (1.3~mm). 
%%%%%%%%
%The (peak) flux changes and of the C3 component in 3C84 during
%the period from late 2015 to the end of 2018 are summarized here.
%%%%%%%%
\textbf{a,}
The light curves of 3C84 at 1.3~mm with SMA
is presented.
\textbf{b,}
The light curve of the  C3 component  
observed at 7mm with KaVA and VLBA  is shown.
\textbf{c,}
The light curve of the  C3 component 
observed at 13mm with VERA and KaVA is presented.
Some error bars cannot be seen because they are smaller than each symbol.
In the panels a, b, and c, 
the black, red, and blue dots are the data for the period before, during, and after the frustrated phase (breakout), respectively.
%%%%%%%%%%%%%%%%%%%%%%%%%%
We show the peak flux densities at 22~GHz and 43~GHz convolved with a 1.0~mas and 0.75~mas circular Gaussian beams, respectively.
While it is clear that the flux significantly
decreased during the frustrated period in all 
three frequency bands, 
it is  also evident that the dimming stopped after the breakout.
%%%%
\textbf{d,}
Time variation of the spectral index of C3
between 22 and 43GHz $\alpha_{22-43}$.
We derived $\alpha_{22-43}$ from 
the peak flux densities convolved with 1~mas at 22~GHz and 43~GHz quasi-simultaneously
acquired within a period of 3 days.
%%%
%During the frustrated phase, it showed the inverted spectrum.
%%%%
%Fermi/LAT's GeV $\gamma$-ray 
%flux above 200 MeV averaged over 3-day intervals as
%measured by Fermi-LAT from photons that passed the diffuse 
%event selection. 
%%%%%
Right: 
\textbf{e,}
The top right panel shows the $n_{\rm cloud}$ dependence 
of the model predicted synchrotron spectra in arbitrary unit.
It can be seen that the spectrum on the low frequency side is more strongly affected by FFA absorption as $n_{\rm cloud}$ increases.
\textbf{f,}
The bottom right panel shows  the corresponding $n_{\rm cloud}$ dependence of the spectral indices for each frequency.
When the density increases to 
$n_{\rm cloud}\approx (4-6)\times 10^{5}~{\rm cm^{-3}}$ 
the positive spectral index can be realized.
\label{fig:LC}}
\end{center}
\end{figure}

In the left panel of Figure~\ref{fig:LC}, 
we show the light curves of the C3 component during 2016-2018.
%%%
As will be explained below, 
the light curve data also agrees well with the prediction of 
the jet-cloud collision.
%%%
%In this figure, we show the  
%evolution of the peak flux of C3 with KaVA and VLBA at 43~GHz
%and the giga electron volt (GeV) $\gamma$-ray flux measured by
%the Large Area Telescope (LAT) on 
%board the recently launched Fermi Gamma-ray Space Telescope.
%%
The radio flux densities at 22~GHz, 43~GHz and 230~GHz deceased
during the frustrated phase (red-colored dots).
These decreases stopped when they entered the breakeout phase
(blue-colored dots).
The synchronization of the period of frustration with that of flux density decrease (red-colored dots) among 22~GHz, 43~GHz and 230~GHz 
suggests an absorption phenomenon.
%%%
Since the hotspot is expected to be buried in the cloud during the collision, the radio emission from the hotspot is considered to be  
partially absorbed by the cloud.
The faster decrease of the radio flux at 22~GHz than that of 43~GHz
supports the free-free absorption (FFA) absorption
that is seen against the counter radio lobe component  
\citep[e.g.,][]{Walker00, Fujita17, Wajima20}
(see N1 in Figure~\ref{fig:overall}).
%%%%%
%There are two absorption mechanisms responsible. 
%One is the free-free absorption (FFA), while the other is 
%synchrotron self absorption (SSA).
%While SSA was in agreement with the magnetic field calculated 
%assuming minimum pressure in the compact radio lobes %\citep[e.g.,][]{Snellen00},
%the FFA absorption is more effective when the synchrotron 
%emitting sources are embedded in dense external environment 
%\citep[e.g.,][]{Walker00, Wajima20}.
%While FFA can naturally explain the link between 
%the frustration and the flux decrease, SSA cannot do it.
%Hence, we adopt FFA hereafter.
%%%%%
Since the degree of FFA depends on the number density of the cloud 
($n_{\rm cloud}$), we can constrain on $n_{\rm cloud}$.
In the past, the C3 component had a optically-thin 
spectrum between 22-43~GHz ($\alpha_{22-43}<0$), which showed
$\alpha_{22-43} \sim -0.9$ in 2008 \citep[e.g.,][]{Suzuki12}.
During the frustrated phase, 
it increased and it maximally reached to $\alpha_{22-43}\sim 0.5$.
Below, we constrain on $n_{\rm cloud}$ based on this $\alpha_{22-43}$ value.
%%%%%
%but it never become a highly inverted spectrum.
%Therefore, $\nu_{\rm FFA}$ is expected be around the 
%range between 22 and 43~GHz.
%%%%%%%%%%%%%%%%%%%%%%%%%%%%%%%%%%%%%
%%%%%%%%%%%%%%%%%%%%%%%%%%%%%%%%%%%%%%
%On the other hand, the decrease of the synchrotron flux density
%at 230~GHz can be probably attributed by an adiabatic expansion 
%loss due to an enlargement the effective size of the 
%decaying jet head C3 indeed seen in Figure~\ref{fig:breakout},
%since FFA never affects the synchtorton spectrum at 230~GHz\ref{fig:FFA}.
%Note that the adiabatic loss may have contributed to the 
%flux decrease at 22 and 43~GHz.
%%
%In the extended data figure~\ref{fig:spectrum-schematic} we show a schematic picture of the 
%radio spectrum behavior that can explain the observed properties
%although a detailed modeling is beyond the scope of this paper.
%%%
It is known that the FFA opacity is given by 
$\tau_{\rm FFA}(\nu) \approx 2.7
\left(\frac{T_{\rm cloud}}{10^{4}~{\rm K}}\right)^{-3/2}
\left(\frac{n_{\rm cloud}}{10^{5}~{\rm cm^{-3}}}\right)^{-2}
\left(\frac{l_{\rm cloud}}{0.1~{\rm pc}}\right)^{1}
\left(\frac{\nu}{43~{\rm GHz}}\right)^{-2}$
where $T_{\rm cloud}$, and
$l_{\rm cloud}$ are the temperature and the path-length
of the dense cloud, respectively
\citep[e.g.,][]{Levinson95, Fujita17}.
%%%%
The observed synchrotron flux density becomes
smaller than the intrinsic flux density by the factor of 
$\exp\left[-\tau_{\rm FFA}(\nu)\right]$.
%%%%%%%%%%%%%%%%%%%%%%%%%%%%%%%%%%%%%
In the right panel of Figure~\ref{fig:LC},
we show the $n_{\rm cloud}$ dependence of the FFA synchrotron flux densities.
To realize the inverted spectra with
$0 \lesssim \alpha_{22-43} \lesssim 0.5$
(light-blue shaded range in the panel $d$), 
$n_{\rm cloud} \sim (4-6) \times 10^{5}~{\rm cm^{-3}}$
(or even higher one) is required.
The estimated $n_{\rm cloud}$ is somewhat
larger than the values of the dynamically estimated density
\citep{Nagai17,Kino18} and the density
estimated by absorption lines detected by ALMA \citep{Nagai19}.
It may indicate that the cloud has 
an internal structure and a denser portion may be responsible 
for the flattening of $\alpha_{22-43}$.
Another intriguing feature seen in Figure~\ref{fig:LC}
is the rise of the spectral index reaching to $\alpha_{22-43}\approx 0$ in mid-2018 again. 
This may be caused by the FFA due to another foreground 
dense cloud floating around C3.

\subsection{Morphological transition of the 
radio lobe after the collision}\label{sec:collapsing}
%%%%%%%%%%%%%
% Fig 3 obs
%%%%%%%%%%%%%
In Figure~\ref{fig:sequence}, 
we have shown the sequence of VLBA images of 3C84 at 43GHz
from 2012 July to 2020 January.
%%%
The emergence of this C3 was around 2003, which means that C3 is only about 15 years old and it has been well known that the shape of C3 is similar to that of the so-called FR II class radio lobes \citep{Nagai10}.
As shown in the top row of Figure~\ref{fig:sequence}, 
C3 continued to propagate southward by mid-2016.
As is in the top row of Figure~\ref{fig:sequence}, 
during the mid-2016 to late 2017 period, the hotspot of the luminosity peak within C3 showed a frustrating trajectory of motion (see Figure~\ref{fig:C3trajectory}) and a corresponding decrease in luminosity (see Figure~\ref{fig:LC}).
%%%
The most important finding in the Figure~\ref{fig:sequence}
is the morphological transition 
of the overall structure of C3 after the breakout.
A series of images taken after 2018 clearly show that the shape of the radio lobe has distorted, losing the characteristics of FR II class radio lobes and changing to have the characteristics of the so-called FR I class radio lobe without a clear hotspot feature.
Such a rapid change of the radio lobe morphology
have not been yet recognized. 
%%%
%and this is the first detection for it.
%%%
What should be emphasized here is that the jet 
received energy from the cloud and strongly influenced its evolution.
This is opposite to the so-called radio-mode feedback where the energy is fed back from jet to the surrounding matter
\citep[e.g.,][]{Fabian12}.

%%%
As is also shown in Figure~\ref{fig:sequence},
we discovered another interesting feature: i.e.,
a new component moving forward in C3 (we denote it as FW).
Interestingly, such behaviors are sometimes seen
in numerical simulations of jet-cloud interactions \citep[e.g.,][]{DalPino99,Wagner12} and thus
the FW might be understood as a circumventing 
portion of the propagating jet during the collision.
%%%

%%%%%
%A study of 3D simulation of jet-dense cloud interactions
%indicates that deflection angle tends to decrease with time 
%as the beam partially penetrates the cloud and describes 
%a C-shaped 
%trajectory around the curved jet-dense cloud contact discontinuity
%when the jet-dense cloud interaction is an off-axis (a.k.a. eccentric) collision.
%\citep{DalPino99}. 
%%%%%%

%If this is the case, then the C3 component is  a long-lived one.
%To realize this scenario, porous and/or inhomogeneity
%in the dense cloud are required \cite{Wagner12}. 
%%%
%Although jet-dense cloud interactions have been theoretically 
%investigated at a galactic nucleus region\citep{Wagner12}, 
%detailed process of vortex generations 
%(and subsequent process of turbulence generations) 
%triggered by jet-dense cloud interactions has not yet been 
%well explored observationally.
%%%%

\section{Discussions}\label{sec:discussion}

%%%%%%
\subsection{Possible Origin of the dense cloud}
%%%%%%

%First, we briefly discuss possible origins of the dense cloud.
%since it is surprising that a dense dense cloud exists 
%in the region on the jet axis, where little material 
%is thought to exist \citep{Ramos-Almeida17}.
%%
%As discussed above our finding of the jet-dense cloud collisions 
%suggests that the dense cloud density on one parsec scale region 
%is much denser than previously thought.
%%%
Since matter cannot maintain a balance of forces on the axis 
because there is no centrifugal force acting on ambient matter,
the region along the jet axis is considered to be essentially free of gas clouds \citep{Ramos-Almeida17}.
Then, how can the cloud we have discovered exist in that region?
%%%
One possible scenario might be a failed clumpy
wind ejected by a different mechanism.
A recent study proposes that
radiation feedback drives a "fountain," that is, a vertical circulation of clumpy gas \citep[][]{Wada12}. 
Since this is not a centrifugal-driven wind but radiation-driven one, the driven gas fountain can occasionally come into the polar region.
%%%%%%%%
%The ratio of ultra-violet luminosity to the Eddington
%luminosity of 3C84 indeed falls in a parameter range 
%in which radiation-driven fountain can work.
%%%%%%%
%Interestingly, this seems to be supported by the 
%recent discovery of the existence
%of polar dust seen in Infra-red band.
%%%%
Another possible scenario
might be a cold chaotic accretion (CCA) 
onto supermassive black holes \citep[][]{Gaspari13}.
In the CCA scenario, 
%the non-linear growth of thermal instabilities
%leads to the condensation of cold clouds and filaments. 
%%%%
%When formation of the multi-phase medium is violent, 
%the mode of accretion is fully cold and chaotic. 
%%%
subsequent recurrent collisions and tidal forces between clouds, filaments promote angular momentum cancellation, 
and it  boosts accretion onto the central black hole.
On sub-pc scales the clouds are channeled to the 
center via a funnel region. 
%%%
Perseus cluster is an archetypal example of a 
cool-core cluster containing plenty of cold gas 
about $10^{10}M_{\odot}$ \citep[][]{salome06}
and the observed complex morphology of the CO filaments 
\citep[e.g.,][]{lim08, Nagai19} on kpc scale
seems to agree with the CCA scenario. 
H$\alpha$ images of the central region of NGC1275
show features in the stellar body of NGC 1275 and 
identify outer stellar regions containing blue, 
probably young, 
star clusters, which can be also 
interpreted as recent accretion of a gas-rich system
\citep[][]{conselice01}, which may support CCA scenario.
%%
%The number density of clouds estimated from our observations
%might place new limits on the above mentioned 
%theoretical models.

%%%
\subsection{New pathway for FR I/FR II divide}
%%%

%As shown in \S~\ref{sec:results}, our VLBI data revealed 
%the existence of a feedback to the radio jet by gas clouds 
%that exists
%in the direction along the jet axis. 
%This is the opposite of the well-known radio-mode feedback 
%from the jet to the surrounding material. 
%This is the new finding of this observation. 
%It would affect the evolutionary track of radio jets.

As mentioned in the Introduction,
two scenarios have been proposed
for the origin of FR~I/FR~II divide: 
(i) it is caused by intrinsic properties 
of a central engine that drives a jet, and 
(ii) it is caused by jet deceleration 
by surrounding ambient matter.
%%%%%%%%%%%%%%%%%%%%%%%%%%%%%%%%%
%The strong jet-clump collision on parsec scale presented in this work would generally affect FR I/II divide.
%%%%%%%%
Within the scenario (ii),
our observational finding adds a new pathway where 
the progenitor of a FR~II radio lobe can turn into the FR~I one 
when it strongly collides with a dense cloud on parsec scale.
Although it is quite interesting to see how often 
such transitions occur via strong deceleration, 
this is the first example and little is known about it so far.
Observational studies of cold gas properties 
at central parsec scale of radio galaxies using ALMA 
would be useful in the future.
%%%%%
%Thus, our observational data of C3 newly shed light on 
%a new process that has been overlooked. 
%%%%%%%%%%%%%%%%%%%%%%%%%%%%%%%

The evolution model where compact radio lobes 
expand and become large FR Is and FR IIs 
are well consistent with radio observational data \citep[e.g.,][]{Kawakatu08, Kunert-B10, O'Dea20}.
\citet[][]{Kawakatu08}
studied the properties of 
more than 100 radio lobes and
by comparing the hot spot advancing speed with the sound speed of the ambient medium, 
they indicated that only compact symmetric objects 
whose initial advance speed is faster than 
about $0.1~c$ can evolve into FR II class. 
Based on the momentum balance between the
jet thrust and the ram pressure by the surrounding ambient matter,
\citet[][]{kawakatu09}
further derived a criterion value of
$L_{j}/n_{\rm amb}\approx 10^{44-45}~{\rm erg~s^{-1}~cm^{3}}$
that divides the evolutionary path to FRI- and FRII-class.
Applying this criterion, 
we find that $n_{\rm amb}=n_{\rm cloud}\sim (4-6) \times 10^{5}~{\rm cm^{-3}}$ can 
easily make the jet frustrated.
The obtained $n_{\rm cloud}$ is roughly comparable to 
the one that caused the jet flip \citep[][]{Kino18}, 
so what could have caused such difference?
It could be interpreted that the flip event was 
an off-axis collision with a small single cloud, 
while the frustration event was a head-on collision 
with a sort of multiple clouds assembly.

%%%%%%%%%%%%%%
\footnote{
On the other hand, when a jet thrust is stronger than a ram pressure from surrounding clouds, radio-mode feedback would be in action as was indicated
in a young radio jet source 4C12.50
\citep[e.g.,][]{Morganti13}.
%the absorption feature and a systemic velocity of atomic hydrogen (HI) and the associated radio jet and lobe image 
%suggested that the radio jet interacts with clouds and made a kinetic-push against the clouds 
}
%%%%%%%%%%%%%%%%%%%%%%%%%%

\begin{figure}[t]
\begin{center}
\includegraphics[trim=0mm 0mm 0mm 0mm, clip, width = 180mm]{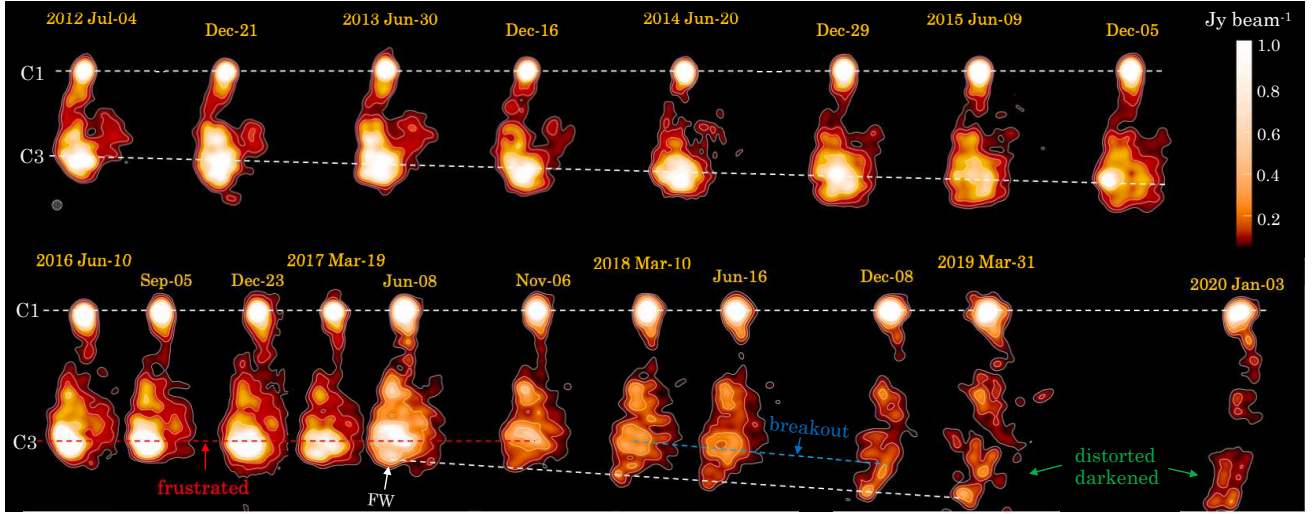}
\caption{\small
Sequence of the archival VLBA images of the 3C84 jet at 43~GHz during the period from 2012 July to 2020 January
(https://www.bu.edu/blazars/). 
%%%%%%%%
All images are restored with the circular Gaussian beam size of 0.25 mas as shown in the bottom-left side of the first image obtained on 2012 July 04.
%%%%%%%%
As shown in the top row, 
C3 continued to propagate southward by mid-2016 (white dashed line).
As is in the bottom row, 
during the mid-2016 to late 2017 period, the hotspot of the luminosity peak within C3 showed a frustrating trajectory of motion (red dashed line).
%%%
Surprisingly, appearance of the radio lobe significantly changes after the breakout. A series of images shows that the shape of the radio lobe has distorted, losing the characteristics of FR II class radio lobes and changing to have the characteristics of the so-called FR I class radio lobes.
%%%%%%%%%%%%%%%%
The new component named as FW has emerged during the 
frustrated phase (white dashed line in the bottom row).
After 2019, the radio lobe 
significantly distorted and darkened.
%%%%%%%
\label{fig:sequence}}
\end{center}
\end{figure}

Since the cloud is of finite size, the jet breaks out of the cloud
in a finite time. 
After the jet breakout, the deposited $E_{\rm feedback}$ would be spent 
for energy-driven expansion of the radio lobes.
By approximating the expansion as
$E_{\rm feedback}$-driven Sedov-Taylor one \citep[e.g.,][]{ostriker88}, 
the predicted expansion velocity can be estimated as
%%%
$v_{\rm exp} 
\approx 0.2c~
\left(\frac{E_{\rm feedback}}{10^{51}~{\rm erg}}\right)^{1/5}
\left(\frac{n_{\rm cloud}}{10^{5}~{\rm cm^{-3}}}\right)^{-1/5}
\left(\frac{t}{{\rm 1~yr}}\right)^{-3/5}$
%%%
that shows a reasonable agreement 
with the C3 advancing speed seen
in Figure \ref{fig:sequence}.
%%%%%%%
\footnote{
The advancing speed of C3 with 5~mas averaged
over 17 years corresponds to the apparent speed as
$\beta_{\rm app}\sim 0.33$.
Together with the jet viewing angle 
$18^{\circ}\lesssim \theta_{\rm view}\lesssim 40^{\circ}$,
a relatively fast speed of 
$0.35 \lesssim \beta\lesssim 0.53$
is estimated.}
%%%%%%%
Further VLBI monitoring observations of 3C84 in the future will 
update this rough-cut order estimation of $v_{\rm exp}$.

What would it be expected
if such a strong a jet-cloud collision generally happens 
and influences of FR I/II divide?
%%%
Then, FR I/II divide would reflect 
information on the dense gas along the jet axis.
For instance, the archetypal FR II class Cygnus A may 
be free from strong collisions with cold dense gas
at its central region, 
since Cygnus A has a classical double radio lobes.
The central cooling time of the Cygnus A cluster 
is about 10 times longer than Perseus cluster \citep[][]{fabian94}
and the longer cooling time may realize CCA free
environment.
On the other hand, 
the Perseus cluster indeed contains filamentary cold molecular gas at its center \citep[e.g.,][]{salome06,lim08}.
It is also known that
there are multiple radio bubbles 
on various directions \citep[][]{pedlar90}.
One may guess that such radio bubbles could be remnants whose
progenitors are young bubbles that underwent strong CCA-jet collisions.
 We further point out the case of 4C31.04, a source with an 
 FR~II morphology on parsec scales and a luminosity that is predicted to evolve in an FR~I large size object \citep[][]{giroletti03}: similarly to 3C84, the hot-spots of 4C31.04 are known to produce feedback on the surrounding gas \citep[][]{zovaro19}.  
 However, differently from 3C84, there is no direct evidence for any morphological transition in that source, perhaps due to the lack of high cadence observations or perhaps because of a different mechanism, not involving jet-CCA interactions.

%%%%%
%Our finding of the changing-look in 3C84 suggests the third scenario for the cause of FR I/FR II divide, i.e., the strong collision between the jet and dense cloud at the central one-parsec scale.  Our result also verifies that the radio-mode feedback occurs not only on galaxy-scale but also on parsec scale.

Asymmetries in young radio lobes also suggest interactions between radio jets and environment \citep{Saikia03}.
Interestingly, 
it is found that a fairly large fraction ($\sim 50$~\%) of 
asymmetric compact radio lobes have the brighter lobes 
closer to the nucleus that can be understood as
jet-cloud interactions \citep[][]{Dallacasa13, Orienti16}.
%%%%%%%%
%When a jet head interacts with a dense cloud its 
%propagation becomes slower than the other radio lobe, and 
%its luminosity increases, reproducing such asymmetries.
%%%%%%%%%
It is curious that a hotspot can be both brightened 
and diminished by a collision. 
It may imply that behavior of the radio flux densities 
of those hotspots highly depends on 
positional relationships between the jet and clouds.
In some cases, clouds may cover up the hotspot like the 
present case of 3C84
%\citep[see also for 4C31.04][]{giroletti03,zovaro19},
while in other cases, they may be exposed and without
being obscured \citep[][]{Dallacasa13, Orienti16}.
Aside from the need for further researches on 
the curious behavior of radio flux densities, 
the existence of young asymmetric radio lobes also suggests 
that jet-cloud interaction on parsec scale
is ubiquitous.

\bigskip
\leftline{\bf \large Acknowledgment}
\medskip

\noindent

This work is mainly based on KaVA observations, which is operated by the Korea Astronomy and Space Science
Institute (KASI) and the National Astronomical Observatory of Japan (NAOJ).
%%%
This work is partially supported by 
JSPS/MEXT KAKENHI (grants 
JP18K03656, JP18H03721, JP18K03709, JP21H01137, 19K03918)
and
by the Korea’s National Research Council of Science 
and Technology (NST) granted by the International joint research project (EU-16-001).
%%%
We thank A. Hirano for her cooperation of VERA 22GHz data analysis.
This study makes use of 43~GHz VLBA data from the VLBA-BU Blazar Monitoring Program (VLBA-BU-BLAZAR;
http://www.bu.edu/blazars/), funded by NASA through the Fermi Guest Investigator Program. 
VLBA is an instrument of the Long Baseline Observatory. The Long Baseline Observatory is a facility 
of the National Science Foundation operated by Associated Universities, Inc.
%%%%%
The Submillimeter Array is a joint project between the Smithsonian Astrophysical Observatory 
and the Academia Sinica Institute of Astronomy and Astrophysics and is funded by the Smithsonian Institution and the Academia Sinica.

%%%%%%%%
\newpage
%%%%%%%%

%%%%%%%%%%%%%%%%%%%%%%%%%%%%%%%%%%%%%%%%%
%%%%%%%%%%%%%%%%%%%%%%%%%%%%%

%%%%%%%%%%%%%%%%%%%%%%%%%
%%%%%%%%%%%%%%%%%%%%%%%%%%

%%%%%%%%%%%%%%%%%%%%%%%%%%%%%%%%%%%%%%%%%%%
%%%%%%%%%%%%%%%%%%%%%%%%%%%%%%%%%%%%%%%%%%%%%%%%%%%%%%%%%%%%%%%%%%

%%%%%%%%%%%%%%%%%%%%%%%%%%%%%%%%%%%%%%%%%%%%%%%%%%%%%%%%%%%%%%%%%%%%%%%%%%%%%%%
%%%%%%%%%%%%%%%%%%%%%%%%%%%%%%%%%%%%%%%%%%%%%%%%%%%%%%%%%%%%%%%%%%%%%%%%%%%%%%%

\bibliography{jetcollision.bib}{}

\end{document}